\documentstyle[twocolumn,aps,epsfig]{revtex}  
\begin{document}
\draft
\title{Quantum Measurements, Information and Entropy Production}
\author{Y.N. Srivastava }
\address{Dipartmento di Fisica \& INFN, Universita' di Perugia, 
I-06100 Perugia, Italy}
\author{G. Vitiello}
\address{Dipartmento di Fisica \& INFM, Universita' di Salerno, 
I-84100 Salerno, Italy}
\author{A. Widom}
\address{Physics Department, Northeastern University, Boston MA 02115 U.S.A.}

\maketitle

\begin{abstract}
In order to understand the Landau-Lifshitz conjecture on the relationship 
between quantum measurements and the thermodynamic second law, we 
discuss the notion of ``diabatic'' and ``adiabatic'' forces exerted by 
the quantum object on the classical measurement apparatus. The notion 
of heat and work in measurements is made manifest in this approach, and 
the relationship between information entropy and thermodynamic entropy is 
explored. 
\end{abstract}  


\narrowtext
\section{Introduction}

Thermodynamics has withstood the onslaught of the two 
revolutions in physical theory that have occurred during this century; i.e. 
relativity and quantum mechanics. Yet the microscopic basis of  
of thermodynamic laws remains unclear. What appears to be time reversal 
symmetry on the microscopic scale turns into time reversal asymmetry 
on the macroscopic scale. Our entire intuition concerning what constitutes 
the ``forward direction in time'' is based upon our observed experience of 
the second law ``increase of entropy''. Yet the foundations of this  
thermodynamic break with reversible time are at best poorly understood. 

Notions of ``quantum measurements'' are {\it also} not very well 
understood. The working physicist usually does not wish to think too hard 
about such a basic problem. For guidance one consults (and repeats) the 
rules of Bohr. The rules are as follows: (i) A measurement is an interaction 
between a classical apparatus and a quantum object. (ii) All data are 
classical and thereby reside in the motion of a classical apparatus; 
and (iii) One may never look directly at a quantum object. 
If one asks experimentally what the quantum object is really doing, then 
the quantum object behaves in a classical fashion as an apparatus. 
If one asks theoretically what the quantum object is doing, then one risks  
being called a philosopher (or perhaps even worse). Bohr\cite{1} 
forbid Einstein to ask what the quantum object was really doing, 
and no agreed upon answer
has yet been formulated. Designing a great experiment using the Bohr rules 
is similar to creating a great sculpture employing a Lagrangian. The real 
art is in making those judgments that are not so well described by dry 
logical formal rules. 

Landau and Lifshitz\cite{2}\cite{3}, in their classical treatises on 
theoretical physics, attempted to connect the above questions concerning 
the nature of the second 
law of thermodynamics and the quantum picture of measurements; They 
conjectured that second law entropy increases are required for quantum 
measurements, and (conversely) that quantum measurements are responsible 
for the second law increase in entropy. A consequence of this conjecture 
is that the second law has no foundation in classical statistical 
thermodynamics. Quantum mechanics must be essential. However, Landau 
and Lifshitz (by their own evaluation) never really proved their 
conjecture. Alas, nor shall we. However, our purpose is to describe (in 
formal detail) the problems as they exist. 

In Sec.2, the classical apparatus Lagrangian is defined. In Sec.3, 
we describe the interaction of this classical apparatus with the 
quantum object. We also introduce the notion of ``diabatic'' and 
``adiabatic'' forces exerted by the quantum object on the classical 
apparatus. In Sec.4 the notion of quantum measurements via the apparatus 
Lagrangian renormalization is discussed. In Sec.5, the formalism is 
illustrated for the Stern-Gerlach experiment\cite{4}. In this experiment, the 
data was sent\cite{5} (in picture post card form) to 
Bohr with a message of 
congratulations for understanding quantum mechanics. Actually, what 
goes on in this experiment is (to this day) somewhat mysterious. 

In Sec. 6, the notion of ``work'' and ``heat'' contributions for the 
energy changes in a measurement ensemble of quantum objects is 
explored in terms of adiabatic forces and diabatic (non-adiabatic) 
transition forces. For macroscopic bodies in the microcanonical 
ensemble, the quantum measurement decomposition of forces into those 
that perform work and those that change energies via heat production 
are shown to be equivalent to the thermodynamic decomposition of 
energy changes into heat and work. The proof, employing the microcanonical 
assumption, is provided in Sec.7. However, we make no claim to have 
derived the thermodynamic laws and their time reversal symmetry breaking 
quality. The situation is what it appeared to be about a century ago. 
The statistical ensembles (microcanonical or canonical) of Gibbs are 
eminently reasonable descriptions of macroscopic systems. But no rigorous 
derivation from dynamics yet exists. 

In Sec.8 we discuss linear friction on the measurement apparatus induced by 
the quantum object. That no such friction term exists if one employs 
the projection postulate is noted in Sec.9. In Sec.10, the information 
entropy is introduced within the context of projective measurements. 
The relationship between information entropy and thermodynamic entropy 
is explored. In the concluding Sec.11, we assert that for quantum objects  
with an infinite number of degrees of freedom it is reasonable to invoke 
a non-unitary time development during the measurement process. 

\section{The Classical Apparatus}

From a classical viewpoint, we describe the classical apparatus motion 
as a path on a manifold with local coordinates 
\begin{equation}
x=(x^1,...,x^n).
\end{equation}
The path $x(t)$ of the apparatus motion, were the apparatus not to 
interact with the quantum object, is determined by the Lagrangian 
\begin{equation}
L_A(x,v)={1\over 2}\mu_{jk}(x)v^jv^k-V(x),
\end{equation}
where the velocity tangent vector $v=(v^1,...,v^n)$ is defined by 
\begin{equation}
v^k=(dx^k/dt),
\end{equation}
and we employ the convention of implicit summation over twice 
repeated (covariant and contravariant) indices. If the apparatus does not 
interact with a quantum object, then the equations of motion read  
\begin{equation} 
{d\over dt}\Big({\partial L_A\over \partial v^k}\Big)=
\Big({\partial L_A\over \partial x^k}\Big).
\end{equation}
Quantum measurements are performed when the quantum object causes the 
apparatus to deviate from the above unperturbed classical paths. 
Such deviations may be described by a renormalization of the apparatus 
Lagrangian induced by an interaction with a quantum object.

\section{Quantum Object} 

The Hamiltonian of the quantum object will be viewed as an operator 
whose matrix elements depend on the apparatus coordinates, 
\begin{equation}
H(x)=\pmatrix
{ 
H_{11}(x) & H_{12}(x) & ... & H_{1m}(x) \cr
H_{21}(x) & H_{22}(x) & ... & H_{2m}(x) \cr
... & ... & ... & ... \cr 
H_{m1}(x) & H_{m2}(x) & ... & H_{mm}(x)  
}.
\end{equation}

If the apparatus path $x(t)$ is given, then the quantum object is thought 
to be described by the Schr\"odinger equation 
\begin{equation}
i\hbar {\partial \over \partial t}\big|\psi (t)\big>=
H\big(x(t)\big)\big|\psi (t)\big>.
\end{equation}
There is no conceptual problem about how a classical apparatus will change 
the state of a quantum object. The manner in which a quantum object changes 
the path of $x(t)$ is a bit more subtle. 

The quantum object interacts with the classical apparatus in virtue 
of the variation of $H(x)$ with $x$; i.e. the force exerted on the 
apparatus by the quantum object is given by 
\begin{equation}
{\cal F}_k(x)=-\Big({\partial H(x)\over \partial x^k}\Big).
\end{equation}
The force is evidently a quantum operator on the quantum system which 
is supposed to have an effect on the classical path $x(t)$. How this 
occurs is not at once obvious. 

It is here convenient to invoke a unitary operator 
\begin{equation}
U^\dagger(x)=U^{-1}(x)
\end{equation}
which obeys 
\begin{equation}
W(x)=U^\dagger (x)H(x)U(x)
\end{equation}  
with $W(x)$ diagonal, 
\begin{equation}
W(x)=\pmatrix
{ 
W_1(x) & 0 & ... & 0 \cr
0 & W_{2}(x) & ... & 0 \cr
... & ... & ... & ... \cr 
0 & 0 & ... & W_{m}(x)  
}.
\end{equation}
From Eqs.(6), (9) and 
\begin{equation}
\big|\psi (t)\big>=U\big(x(t)\big)\big|\Psi (t)\big>
\end{equation}
it follows that 
\begin{equation}
i\hbar {\partial \over \partial t}\big|\Psi (t)\big>=
{\cal H}\big(x(t),v(t)\big)\big|\Psi (t)\big>,
\end{equation}
where 
\begin{equation}
{\cal H}(x,v)=W(x)-v^kP_k(x),
\end{equation}
and 
\begin{equation}
P_k(x)=i\hbar U^\dagger (x){\partial U(x)\over \partial x^k}=
-i\hbar {\partial U^\dagger (x)\over \partial x^k}U(x).
\end{equation}
\bigskip

From Eqs.(12) and (13), it follows that ${\cal H}(x,v)$ is the 
quantum object Hamiltonian when the interaction is with an apparatus 
whose motion exhibits coordinates $x$ and velocities $v$. The force 
operator acting on the apparatus follows from Eqs.(9) and (14) to be 
\begin{equation}
U^\dagger (x) {\cal F}_k(x)U(x)=F_k(x)+f_k(x)
\end{equation}
where 
\begin{equation}
F_k(x)=-\Big({\partial W(x)\over \partial x^k}\Big)
\end{equation}
is the ``adiabatic'' part of the force, and  
\begin{equation}
f_k(x)=-{i\over \hbar }\big[W(x),P_k(x)\big]
\end{equation}
is the ``diabatic'', i.e. ``non-adiabatic'', part of the force.

\section{Adiabatic Measurement Lagrangians}

The object Hamiltonian in Eq.(5) allows for the notion of a quantum 
measurement as a renormalization of the apparatus Lagrangian in Eq.(2). 
Explicitly, if one defines a measurement basis for the quantum object 
employing the adiabatic energy levels of Eqs.(9) and (10), 
\begin{equation}
H(x)|k>=W_k(x)|k>,
\end{equation}
where
\begin{equation}
<k'|k>=\delta_{k',k},
\end{equation} 
then a general wave function may be written in the adiabatic basis  
\begin{equation}
|\psi >=\sum_k C_k |k>.
\end{equation}
The renormalized measurement Lagrangian for the classical apparatus 
may be defined as a mapping from an adiabatic basis state to 
an effective apparatus Lagrangian corresponding to that state; i.e.  
\begin{equation}
|k>\to {\cal L}_{k}(x,v)=L_A(x,v)-W_k(x),
\end{equation}
where $L_A(x,v)$ is defined in Eq.(2). 

Quantum measurements are described as follows: (i) All data are 
classical. For a classical apparatus with coordinates $x$, the 
``data'' are the ``apparatus readings'', i.e. classical paths $x(t)$.
(ii) If the quantum object were in the state $|k>$, then the renormalized 
apparatus Lagrangian ${\cal L}_{k}(x,v)$ would control the classical path 
$x(t)$ via 
\begin{equation} 
|k>\to 
\Big\{
{d\over dt}\Big({\partial {\cal L}_{k}\over \partial v^k}\Big)=
\Big({\partial {\cal L}_{k}\over \partial x^k}\Big)
\Big\}.
\end{equation} 
(iii) The probability that the state $|k>$ is ``observed'' (i.e. that 
the apparatus yields a characteristic reading $x(t)$ from the Lagrangian  
${\cal L}_{k}$) is given by 
\begin{equation}
P_k=|C_k|^2
\end{equation}
where the amplitude $C_k$ is defined in Eq.(20).

Thus, the quantum object Hamiltonian gives rise to a Lagrangian 
renormalization via Eqs.(18) and (21). The measurement process 
assigns to each adiabatic state $|k>$ of the measurement basis an 
apparatus Lagrangian ${\cal L}_k$ yielding a characteristic reading 
$x(t)$ from state $|k>$ with quantum probability $P_k=|C_k|^2$. 
This prescription is in agreement with the Bohr notion of how a quantum 
object yields  a classical apparatus reading. This forms the basis of 
``quantum measurements'' in which all ``data'' are classical in accordance 
with the so-called ``Copenhagen School''. A simple example of the above 
approach to measurements is the Stern-Gerlach experiment. Let us consider 
this in more detail.

\section{The Stern-Gerlach Experiment}

As an example of the above formalism let us consider the Stern-Gerlach 
beam splitting experiment wherein a beam of spin one-half neutral atoms 
with a gyromagnetic ratio $\gamma $ passes through a spatial region 
with an inhomogeneous field ${\bf B}({\bf r})$. The two by two spin 
Hamiltonian matrix corresponding to Eq.(5) reads (for this example) as 
\begin{equation}
H({\bf r})=-\hbar \gamma {\bf S\cdot B}({\bf r}),
\end{equation}
while the eigenstates corresponding in Eq.(18) are determined by 
\begin{equation}
H({\bf r})|\pm >=W_\pm ({\bf r})|\pm >,
\end{equation}
where 
\begin{equation}
W_\pm ({\bf r})=\mp 
\Big({\hbar \gamma |{\bf B}({\bf r})|\over 2}\Big).
\end{equation}
Thus, if the incoming beam is described as in Eq.(20) by the quantum state 
\begin{equation}
|\psi >=C_+|+>+C_-|-> 
\end{equation}
then with probability $P_+=|C_+|^2$ an atom in the beam follows the 
classical path specified by the classical equation of motion 
\begin{equation}
|+>\to \Big\{M{d^2{\bf r}\over dt^2}={\bf F}_+({\bf r})
=-{\bf \nabla }W_+({\bf r})\Big\},
\end{equation}
and with probability $P_-=|C_-|^2$ an atom in the beam follows the 
classical path specified by the classical equation of motion 
\begin{equation}
|->\to \Big\{M{d^2{\bf r}\over dt^2}={\bf F}_-({\bf r})
=-{\bf \nabla }W_-({\bf r})\Big\},
\end{equation}

The Stern-Gerach experiment represents the success of the Bohr 
theoretical interpretation of quantum mechanics at its very best. 
The quantum object is simply the spin ${\bf S}$ of the atom, which 
yields two possible states $|\pm >$. The classical apparatus reading 
is the path ${\bf r}(t)$ of the atom which obeys one of two possible 
classical \{Eqs.(28) or (29)\} of motion. All of this may be described 
by the adiabatic Lagrangians, as in Eq.(21) 
\begin{equation} 
|\pm >\to {\cal L}_\pm ({\bf r},{\bf v})
={1\over 2}M{\bf v}^2-W_\pm({\bf r})
\end{equation}
where $W_\pm({\bf r})$ is defined in Eq.(26). Thus, the triumphant 
Bohr could predict that for a beam described by the quantum state in 
Eq.(27), with probability $|C_\pm|^2$ one would find an atom of the 
beam moving along a classical path with a reading controlled by the 
Lagrangian ${\cal L}_\pm ({\bf r},{\bf v})$. The picture 
post card\cite{5} with the note of congratulations from Gerlach to Bohr 
was at the start of an era of complete faith (if not blind then perhaps 
just near sighted) in quantum mechanics. The Copenhagan interpretation 
of quantum mechanics was perfect!

\section{Work and Heat}

In the Lagrangian description of Sec.4, the effective Lagrangians were 
computed from the adiabatic energy levels $\{W_k(x)\}$. When the external 
coordinates are changed, the resulting induced changes in the adiabatic 
energy levels describe the thermodynamic notion of {\em work} done by the 
classical environment on the quantum object(or the reverse). 

Another method of changing the energy of the quantum object, is by causing 
a transition from one adiabatic energy level to another adiabatic energy 
level. In thermodynamic terms such diabatic transitions describe the 
{\em heat } flow from the classical environment to the quantum object, 
(or the reverse).    

Let us consider in more detail the decomposition of energy changes 
into ``work'' and ``heat''. The mean energy of quantum objects as members 
of a statistical ensemble may be written as 
\begin{equation}
\bar{E}=Tr\big(\rho W\big),
\end{equation}
where $\rho $ is the density matrix of the ensemble which obeys 
\begin{equation}
i\hbar \Big({\partial \rho \over \partial t }\Big)=
[{\cal H},\rho ], 
\end{equation} 
where ${\cal H}$ is defined in Eq.(13).
From Eq.(31)
\begin{equation}
{d \bar{E}\over dt}=Tr\Big({\partial \rho \over \partial t }W\Big)
+Tr\Big(\rho {\partial W\over \partial t}\Big),
\end{equation}
which implies 
\begin{equation}
{d \bar{E}\over dt}=-{i\over \hbar }Tr\big([{ \cal H},\rho ]W\big)
+Tr\Big(\rho {\partial W\over \partial x^k}\Big)v^k,
\end{equation}
where Eq.(3) has been invoked. Note that the first term on the right hand 
side of Eq.(33) may be simplified according to 
\begin{equation}
Tr\big([{\cal H},\rho ]W\big)=
Tr\big(\rho [W,v^kP_k]\big)
\end{equation}
where Eq.(13) has been invoked. Furthermore, from Eqs.(13), (16), (17), (34) 
and (35) it follows that 
\begin{equation}
{d \bar{E}\over dt}=-\big(\bar{f}_k+\bar{F}_k\big)
{dx^k\over dt}.
\end{equation}

Eq.(36) is the central result of this section. The mean energy changes in 
the quantum object (as a member of a statistical ensemble) can be decomposed 
into heat and work; i.e. 
\begin{equation}
d\bar{E}=d^\prime Q+d^\prime W,
\end{equation}
where the work done on the quantum object by the classical environment is 
determined by  
\begin{equation}
d^\prime W=-\bar{F}_kdx^k,
\end{equation}
and where the heat is determined by 
\begin{equation}
d^\prime Q=-\bar{f}_kdx^k.
\end{equation}
We employ the ``inexact differential'' $d^\prime $ notation for the 
heat $d^\prime Q$ (energy changes from quantum transitions between 
adiabatic energy levels) and the work $d^\prime W$ (energy changes 
from deformations in the given adiabatic energy levels). This decomposition 
of the mean energy change is exactly (i.e. rigorously) true, and constitutes 
a general derivation of the first law of thermodynamics.

\section{Statistical Thermodynamics}

The decomposition of energy changes into work and heat in Eqs.(37), (38) 
and (39) did not require that the quantum object be large or complex. 
An ensemble of two level atoms as used in the Stern-Gerlach atomic beam 
experiment is sufficient for the notion of introducing diabatic and 
adiabatic forces. 

For the purpose of describing statistical thermodynamics, as employed for  
systems with a macroscopic number of particles, one normally 
introduces one (canonical or microcanonical) of the Gibbs probability 
distributions. For a macroscopic system, described by an infinite 
matrix $W(x)$, one may define $\Omega (E,x)$ as the number of quantum 
states with energy equal or less than $E$; i.e. with the step function 
\begin{equation}
\vartheta(E)=\cases{1, &if $E\ge 0$;\cr 0, &otherwise, \cr}
\end{equation}
we have 
\begin{equation}
\Omega(E,x)=Tr \vartheta\big(E-W(x)\big).
\end{equation}
The microcanonical entropy of the quantum system is then 
\begin{equation}
S(E,x)=k_B\ ln\Omega (E,x).
\end{equation}
The microcanonical temperature $T$,
\begin{equation}
{1\over T}=\Big({\partial S(E,x)\over \partial E}\Big),
\end{equation}
is related to the microcanonical density of states, 
\begin{equation}
G(E,x)=Tr\delta \big(E-W(x)\big),
\end{equation}
via 
\begin{equation}
\Omega (E,x)=k_BT(E,x)G(E,x).
\end{equation}

In the thermodynamic limit of large quantum objects, the mean adiabatic 
force on the apparatus is given in the microcanonical ensemble by 
\begin{equation}
\bar{F}_k=-\Big({Tr\big(\delta\big(E-W\big)(\partial W/\partial x^k)\big)
\over Tr\delta\big(E-W\big)}\Big),
\end{equation} 
where Eqs.(16) has been invoked. From Eqs.(42)-(46), it follows that 
\begin{equation}
\bar{F}_k=T\Big({\partial S\over \partial x^k}\Big)_E=
\Big({\partial E\over \partial S}\Big)_x
\Big({\partial S\over \partial x^k}\Big)_E,
\end{equation}
or equivalently  
\begin{equation}
\bar{F}_k(x,S)=-\Big({\partial E(x,S)\over \partial x^k}\Big).
\end{equation}
For quasi-static changes in a macroscopic thermodynamic system we 
have in the microcanonical ensemble both the first and second law of 
thermodynamics in the form 
\begin{equation}
dE=TdS-\bar{F}_kdx^k.
\end{equation}

Note, from Eqs.(37), (38), (39), and (49), that the inexact heat 
differential from the diabatic forces has the temperature as an 
integrating factor 
\begin{equation}
dS=\Big({d^\prime Q\over T}\Big)
=-\Big({\bar{f}_kdx^k\over T}\Big),
\end{equation}
as required by the second law implied existence of an entropy function. 
If one makes the Gibbs microcanonical assumption for quasi-static 
(sufficiently slow) changes, then one achieves the thermodynamic 
laws in a consistent manner. 

However we have not thus far justified the Gibbs microcanonical 
assumption by anything deeper than the mere observation that 
this assumption is reasonable. It is here that we fall short of a 
strict mathematical derivation of the thermodynamic second law. 

Our instinct is that one must consider a quantum object with an 
infinite number of degrees of freedom, e.g. as conventionally done 
in quantum field theory. For the field theory case, the states 
corresponding to {\em different} values of $x$ are not unitarily 
equivalent. Thus, there is no unitary transformation of the statistical 
state of the quantum object\cite{6} when one changes the $x$ parameters 
from $x_{inital}$ to $x_{final}\ne x_{inital}$.
While the usual unitary time development involves {\em only} the Hamiltonian, 
the {\em necessarily}  diabatic time development between {\em unitarily 
inequivalent} Hilbert spaces involves {\em also} the entropy, or 
equivalently the free energy.

\section{Friction in a Macroscopic Apparatus}

Let us now return to the apparatus viewed as having contact with a 
macroscopic thermodynamic system. In Eq.(2) we take the apparatus potential 
$V(x)$ to be determined by the apparatus adiabatic energy function 
$E_A(x,S)$; i.e. 
\begin{equation}
L_A(x,v)={1\over 2}\mu_{jk}(x)v^jv^k-E_A(x,S),
\end{equation}
and we may further include the diabatic apparatus friction forces 
$\{\bar{f}_{Ak}\}$ which 
heat the macroscopic degrees of freedom by 
\begin{equation}
{d\over dt}\Big({\partial L_A\over \partial v^k}\Big)-
\Big({\partial L_A\over \partial x^k}\Big)=\bar{f}_{Ak}.
\end{equation}
For small velocities it is often sufficient to take the 
apparatus friction forces to be linear functions of the velocities, 
\begin{equation}
\bar{f}_{Ak}=\Gamma_{kj}(x)v^j.
\end{equation}
Standard linear response theory gives expressions for the friction 
coefficients 
\begin{equation}
\Gamma_{kj}(x)={i\over \hbar}
\int_0^\infty \big<[f_{kA}(x,t),P_{Aj}(x)]\big>dt,
\end{equation}
where Eqs.(13) and (17) has been invoked as well as the Heisenberg 
picture
\begin{equation}
f_{Ak}(x,t)=e^{iW_A(x)t/\hbar}f_{Ak}(x)e^{-iW_A(x)t/\hbar}.
\end{equation}
If one employs the canonical density matrix, 
\begin{equation}
\rho_A ={1\over Z_A}e^{-\beta W_A(x)/\hbar },\ 
\beta=\Big({\hbar\over k_BT}\Big),
\end{equation}
which is equivalent to the microcanonical ensemble (for sufficiently 
large thermal systems), then the expression for the friction coefficients 
reduce to the Nyquist-Kubo formula\cite{7},
$$
\Gamma_{kj}(x)=
$$ 
\begin{equation}
{1\over \hbar}
\int_0^\beta d\lambda  \int_0^\infty dt 
\big<f_{jA}(x,-i\lambda )f_{kA}(x,t)\big>,
\end{equation} 
wherein the friction coefficient is related to the quantum fluctuations 
of the diabatic forces.

Thus in any apparatus one does expect some friction forces as in Eq.(52). 
The total apparatus heating rate 
\begin{equation} 
\Big({d^\prime Q_A\over dt}\Big)=f_{Ak}v^k
\end{equation}
in the linear regimes of Eq.(53) is described by the Rayliegh 
dissipation which is quadratic in the velocities 
\begin{equation}
\Big({d^\prime Q_A\over dt}\Big)=\Gamma_{kj}(x)v^kv^j.
\end{equation}

For the case of the Stern-Gerlach experiment, the apparatus friction 
forces on the atoms with spin are quite small. The dissipation in 
this experiment amounts to the small friction forces induced by 
Ohms law eddy currents inside the deflecting magnets in the beam 
splitter.  

\section{Low Velocity Projective Measurements are Frictionless}

Within the above discussion of quantum measurements in the adiabatic 
representation of Eq.(10), the statistical state of a quantum object 
is described by the matrix 
\begin{equation}
\rho=\pmatrix
{ 
\rho_{11} & \rho_{12} & ... & \rho_{1m} \cr
\rho_{21} & \rho_{22} & ... & \rho_{2m} \cr
... & ... & ... & ... \cr 
\rho_{m1} & \rho_{m2} & ... & \rho_{mm}  
}.
\end{equation}
The unitary time development of the statistical state obeys the equation 
of motion
\begin{equation}
i\hbar \Big({\partial \rho \over \partial t}\Big)=
[{\cal H},\rho ]
\end{equation}
where ${\cal H}$ is given in Eq.(13). However, in the fairly early days of 
quantum mechanics, von Neumann introduced, over and above the standard 
Bohr Copenhagen interpretation, the notion of a projective measurement
\cite{8}. During a projective measurement, the density matrix is thought 
to obey a non-unitary time development governed by 
\begin{equation} 
\rho \to \rho_P =\sum_k P_k\rho P_k,
\end{equation}
for a set of projection operators $P_k=P_k^\dagger =P_k^2$ which is 
complete $\sum_k P_k=1$. For the problem at hand $P_k=|k><k|$, so that 
the projected matrix 
\begin{equation}
\rho\to \rho_P=\pmatrix
{ 
\rho_{11} & 0 & ... & 0 \cr
0 & \rho_{22} & ... & 0 \cr
... & ... & ... & ... \cr 
0 & 0 & ... & \rho_{mm}  
}
\end{equation}
has the above diagonal form.

Note, from Eqs.(10), (17) and (63), it follows that the mean diabatic 
(frictional heating) force on the quantum object obeys 
\begin{equation}
\bar{f}_{Pk}=Tr\big(\rho_P f_k\big)=0.
\end{equation} 
The frictional force is zero, when computed from the von Neumann 
projected density matrix. No diabatic heat is generated in a 
low velocity projective measurement. This strongly 
suggests that projective measurements (in general) violate 
energy conservation, which is indeed true. Suppose that the projection 
operator does not commute with the total energy operator of the 
apparatus plus the quantum object. Further suppose that before the 
measurement the system plus apparatus have a sharply defined energy. 
After the measurement the energy value will no longer be sharp since 
the projection operator does not commute with the total energy. Thus, 
with finite probability, the total energy is not conserved. We consider 
this violation to be a significant flaw in the projective measurement 
picture. Let us reconsider the situation without imposing the von 
Neumann projection postulate. 

In the Heisenberg (unitary time development) picture, the diabatic 
friction force operator obeys  
\begin{equation}
\Big({d f_k\over dt}\Big)=\Big({\partial  f_k\over \partial t}\Big)
+{i\over \hbar }\big[{\cal H},f_k\big].
\end{equation}
With adiabatic Bohr transition frequencies 
\begin{equation}
\hbar \omega_{ij}(x)=W_i(x)-W_j(x),
\end{equation}
and for a slowly moving apparatus (small $v^k$), the diabatic friction 
force operator has only quickly oscillating off diagonal matrix elements 
\begin{equation} 
<j|f_k(t)|i>\sim e^{-i\int \omega_{ij} (x(t))dt}<j|f_k|i>
\end{equation}
whose time average is virtually zero. Again, in a time averaged sense, 
the mean diabatic friction force is negligible, $\bar{f}_k\approx 0 $. 
For the case of a Stern-Gerlach experiment, the time averaged null 
behavior of the rapidly varying diabatic friction force has been 
discussed elsewhere\cite{9}.

Thus, in the general case, the mean energy changes of an ensemble 
of quantum objects decomposes into a ``heat'' and ``work'' term 
\begin{equation}
d\bar{E}=d^\prime Q + d^\prime W 
\end{equation}
with the inequality 
\begin{equation}
|d^\prime Q|<<|d^\prime W|, \ {\rm projective\ measurement.}
\end{equation}

We stress that the diabatic friction forces can be small (on the average) if 
the apparatus velocities are also small. However, even if the diabatic  
forces during a projective measurement on the average are small, they 
are often still present as ``noise''. Via the fluctuation-dissipation 
Eq.(57), such noise does give rise to friction at least for higher 
velocities if the apparatus is (at least in part) macroscopic. In the 
limit of an infinite number of degrees of freedom, one is 
not allowed to totally ignore heating, since (as discussed in 
Sec.7) the diabatic forces are invoked as part of the dynamics of 
passing between two unitarily inequivalent spaces.  

\section{Information and Entropy}

In some treatments of statistical thermodynamics, one associates the 
physical entropy to the statistical ``information'' contained in 
$\rho $\cite{8}\cite{10}\cite{11}; i.e. 
\begin{equation}
{\cal S}=-k_B\ Tr\big(\rho ln \rho \big). 
\end{equation}
The difficulty with this approach is that the information entropy 
${\cal S}$ is not always simply related to thermodynamic heat 
$d^\prime Q=-\bar{f}_kdx^k$. For example, von Neumann has proved\cite{8}
for the projective measurement in Eq.(62), that 
\begin{equation}
{\cal S}_P=-k_B\ Tr\big(\rho_P ln \rho_P \big)
\end{equation}
and that the change in statistical entropy obeys 
\begin{equation} 
\Delta{\cal S}={\cal S}_P-{\cal S}\ge 0.
\end{equation}
Thus the statistical entropy of a projective measurement is thought to 
increase, yet the frictional heating (i.e. {\it thermodynamic} entropy 
increase $d'Q/T$) in a low velocity projective measurement is virtually 
zero.

The paradox may be resolved when it is realized that during the 
{\it unitary time development} of Eq.(61), it is in fact rigorously 
true that the statistical entropy of Eq.(70) does not increase at all; 
i.e.  
\begin{equation}
{d{\cal S}\over dt}=-k_B {d\over dt } Tr\big(\rho ln \rho \big)=0.
\end{equation}
Thus it is only through the projection postulate itself, that the 
statistical entropy may be shown to increase. 

Finally, non-unitary time development can enforce the approximate 
equality of the thermodynamic entropy in Eq.(42) and the statistical 
entropy Eq.(70) as in Sec.7. This is likely to happen in the limit 
of an infinite number of degrees of freedom. It is in such a limit 
that the information theory picture may be properly formulated, also 
with possible advantage in problems of quantum computation\cite{12}.

\section{Conclusions}

 In quantum mechanics, the unitary time development of the statistical 
state $\rho $ is all that we know for sure to be true. 
Under some circumstances, this density matrix $\rho $ may 
be approximated by a projected density matrix $\rho_P$. The reason why 
this replacement may not lead to error, is that there exists very rapid phase 
oscillations which average out (e.g. for the diabatic forces) and which cannot 
later be resolved experimentally.

If such measurements are statistically accurately described by the 
projection postulate, via $\rho \to \rho_P$, then there is a quantity which 
we have called statistical entropy which increases, as was early proved by 
von Neumann. A difficulty with this definition of entropy, via 
information theory, is that the statistical entropy is not {\em always} simply 
related to the thermodynamic notion of heat $d^\prime Q=-\bar{f}_kdx^k$.
For the case of the (macroscopically equivalent) Gibbs 
ensembles, the statistical entropy is closely related to heating 
$d^\prime Q=TdS$. The reasoning applied to the Gibbs ensembles 
is that statistical entropy is at maximum (subject to conservation law 
constraints). Then the statistical 
entropy does indeed obey $d^\prime Q=TdS$, but then there is little 
difference between the statistical entropy and the thermodynamic 
$S=k_B\ ln\Omega$. 

If the quantum object has an infinite number of degrees of freedom,
it is reasonable to invoke a non-unitary time development between 
Hilbert spaces which are not unitarily equivalent. 
For this situation, it appears reasonable that the equivalence between 
the statistical entropy and the thermodynamic entropy may be established 
during irreversible dynamics. 

Although we do not pretend that a complete proof of the 
physical picture has been provided, we feel that 
the discussion presented in this work accurately describes 
the problem of understanding the connection between quantum 
measurements and the second thermodynamic law of entropy increase.  
\medskip 
\par \noindent
This work has been partially supported by the INFN, INFM and MURST.


\begin{thebibliography}{99}
\bibitem{1} A. Fine, {\it The Shaky Game: Einstein, Realism and Quantum 
Theory}, University of Chicago Press, Chicago (1986).
\bibitem{2} L. D. Landau and E. M. Lifshitz, {\it Quantum Mechanics: 
Non-Relativistic Theory}, Pergamon Press, Oxford, (1977).
\bibitem{3} L. D. Landau and E. M. Lifshitz, {\it Statitsical Physics}, 
Pergamon Press, Oxford, (1978).
\bibitem{4} Von W. Gerlach and O. Stern, {\it Z. Phys.} {\bf 8}, 110, 
(1921).
\bibitem{5} J. S. Townsend, {\it A modern Approach to Quantum Mechanics}, 
McGraw Hill Inc., New York, (1992).  
\bibitem{6} M. Blasone, Y. N. Srivastava, G. Vitiello and A. Widom, 
{\it Annals of Phys.} {\bf 267}, 61, (1998). 
\bibitem{7} M. Toda, R. Kubo and N. Saito, {\it Statistical Physics} 
{\bf 2}, Springer Verlag, New York (1991).
\bibitem{8} J. von Neumann, {\it Mathematical Foundations of Quantum 
Mechanics}, Princeton University Press, Princeton (1955).
\bibitem{9} M. Hannout, S. Hoyt, A. Kryowonos and A. Widom, 
{\it Am. J. Phys.} {\bf 66}, 377, (1998).
\bibitem{10} E. T. Jaynes, {\it Phys. Rev.} {\bf 106}, 620 (1957);
{\it Phys. Rev.} {\bf 108}, 171 (1957).
\bibitem{11} H. S. Leff and A. F. Rex (Eds.), {\it Maxwell's Demon: 
Entropy, Information, Computing}, Princeton University Press, Princeton 
(1988).
\bibitem{12} C. P. Williams and S. H . Clearwater {\it Explorations in 
Quantum Computing}, Springer-Verlag, New York (1998).



\end{thebibliography}
\end{document}